\documentclass[twocolumn,prl]{revtex4}
\usepackage{graphicx,amssymb,amsbsy}

\def\avg#1{\langle#1\rangle}	\def\<{\langle} 	\def\>{\rangle}

\def\al{\alpha}				
		\def\Del{\Delta}
\def\eps{\epsilon}

 \def\Vp{{\mathbf p}} \def\Vq{{\mathbf q}} 
\def\VQ{{\mathbf Q}}	
\def\Vx{{\mathbf x}}  \def\V0{{\mathbf 0}}

\def\be{\begin{equation}}	\def\ee{\end{equation}}
\def\bea{\begin{eqnarray}}	\def\eea{\end{eqnarray}}

\begin{document}
\title{Interior Gap Superfluidity}
\author{W. Vincent Liu and Frank Wilczek}
\affiliation{Center for Theoretical Physics, Department of Physics,
Massachusetts Institute of Technology,
Cambridge, Massachusetts 02139}

\preprint{MIT-CTP \# 3279} 

\begin{abstract}

We propose a new state of matter in which the pairing interactions
carve out a gap within the interior of a large Fermi ball, while the
exterior surface remains gapless.  This defines a system which
contains both a superfluid and a normal Fermi liquid simultaneously,
with both gapped and gapless quasiparticle excitations.  This state
can be realized at weak coupling.  We predict that a cold mixture of
two species of fermionic atoms with different mass will exhibit this
state.  For electrons in appropriate solids, it would define a
material that is simultaneously superconducting and metallic.

\end{abstract}

\maketitle


Recent developments in ultracold alkali atomic gases \cite{UCA:nature}
have revitalized interest in some basic qualitative questions of
quantum many-body theory, because they promise to make a wide variety
of conceptually interesting parameter regimes, which might previously
have seemed academic or excessively special, experimentally
accessible. With this motivation, and stimulated by questions in
quantum chromodynamics (QCD) at high density
\cite{Schaefer-Wilczek:99,+Rajagopal:01,+Wilczek:00}, 
we here revisit the question of fermion
pairing between species whose Fermi surfaces do not precisely match.
We have found a possibility that seems to be new and certainly is
interesting, and which could turn out to be relevant even for
conventional solids.

The standard Bardeen-Cooper-Schrieffer (BCS) \cite{BCS:57} theory of
superconductivity describes pairing between particles of equal and
opposite momentum near a common Fermi surface. 
For classic s-wave superconductors the
pairing occurs between electrons of opposite spin.
In the presence of a weak magnetic field, and in particular in the
case of ferromagnetic order, the Fermi surfaces of the opposite spins
will not match, and the Cooper pairing instability, which was enhanced
by vanishing energy denominators, will no longer occur at arbitrarily
weak coupling. Larkin and Ovchinnikov and independently
Fulde and Ferrell \cite{LOFF:65+64} showed that in this
circumstance it might be favorable to effectively relatively translate
the Fermi surfaces, pairing at a non-zero total momentum (LOFF phase).

A simpler situation, conceptually, is that pairing occurs between two
species whose Fermi surfaces do not match simply because their
densities or effective masses differ.  This possibility arises in
several contexts.  (i) In ultracold atom systems, it could occur
simply because there are atoms of different elements.  (ii) In solids
it could occur for electron populations in two different bands.  (iii)
In QCD it occurs for different species of quarks (up, down, strange).
If the mismatch is small and the two species are alternative states of
the same particle (such as the spin up and down states of electrons or
two hyperfine-spin states of cold  $^{40}$K 
 or $^6$Li atoms as prepared in experiments
\cite{+Jin:99,+Jin:01,+Thomas:00,+Thomas:02}),
it can be favorable to equalize the Fermi surfaces, absorbing a cost
in kinetic energy, and then to pair at zero momentum following BCS.
For larger mismatches, LOFF-type ordering can occur.  More elaborate
forms of position-dependent ordering, with the superfluid gap having
standing wave or even crystalline structure, have been found to be
favorable in models of QCD at high density \cite{+Rajagopal:02pre}.

None of these possibilities, however, extrapolates to what we might
expect at strong coupling.  Given strong attraction between the
species, we would just expect to bind as many paired quasi-molecules
as possible.  At low temperature, these quasi-molecules will form a
Bose-Einstein condensate.  The residual unpaired particles will
constitute a separate normal fluid.  It is natural to inquire whether
there is a weak-coupling phase that matches this behavior
qualitatively.  We now identify such a phase.

\begin{figure}[htbp]
\begin{center}
\includegraphics[width=\linewidth]{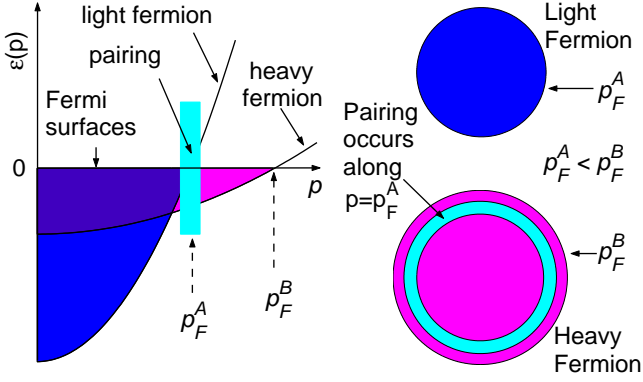}
\end{center}
\caption{The prototype situation where we anticipate formation of an
interior gap superfluid at weak coupling.  There are two species of
fermions with different band structures, here both taken as isotropic
and parabolic, but with different effective masses and different sizes
in momentum space.  When the larger Fermi ball derives from relatively
flat band (large effective mass) and the interaction near the momentum
surface defined by the smaller Fermi sphere is attractive, it can be
favorable to form correlated pairs near this smaller sphere, even at
the cost of promoting some particles of the heavier species to the
exterior Fermi sphere.  One will then have both superfluidity, with a
momentum gap at the smaller sphere, and normal Fermi liquid excitations
at the larger sphere.}
\label{fig:bands}
\end{figure}

Consider a homogeneous fermion gas in three dimensions containing 
two species ($\al=A,B$) obeying
simple parabolic dispersion relations, described by the Hamiltonian 
\begin{equation}
H= \sum_{\Vp\al} \eps^\al_\Vp {\psi^\al_{\Vp}}^\dag {\psi^\al_{\Vp}}
 + g\sum_{\Vp\Vp^\prime\Vq} \psi^\dag_{A \Vq+\Vp}
\psi^\dag_{B \Vq-\Vp} \psi^B_{\Vq-\Vp^\prime}
 \psi^A_{\Vq+\Vp^\prime} 
\label{eq:H}
\end{equation}
where $\epsilon_\al(\Vp) = \Vp^2/2m_\al - \mu_\al$.  
See Fig.~\ref{fig:bands}.  (More precisely, we assume that this
interaction exists so long as the momenta of the particles are within
a strip of size $\lambda$ around the smaller Fermi surface;
$\lambda$ will later serve as an ultraviolet cutoff.)  Our heuristic
analysis will not distinguish whether or not the species are strictly
conserved separately.  We define chemical potentials so that the Fermi
surfaces for both species are at $\eps_F=0$.  We shall be interested
in cases where $m_A<m_B$ and $\mu_A>\mu_B$, in such a way that the
Fermi momentum for the species B is greater than that for the species
A, $p_F^B > {p}_F^A$.

We suppose that there is an attractive effective interaction 
in s-wave between particles of different species ($g<0$),
and that the coupling is weak, so that we can construct our
ground-state by modifying the ground state of the non-interacting system.
If the Fermi surfaces matched, the attractive interaction would trigger
standard BCS superfluidity, with Cooper pairing of equal and opposite
momenta.   The BCS wave function, however,
postulates either zero or double occupancy of the paired modes, and it is
incompatible with keeping the modes of species B between $p_F^A$ and
$p_F^B$ completely filled.  If we are to support pairing of total momentum zero
we must promote some
particles of species B up to momenta near $p_F^B$, thus carving an
interior 
``trench'' of the species B Fermi sea near momentum $p=p_F^A$.

There is competition between the energetic cost of such promotions and
the gain from pair-formation, and it may not be obvious whether there
can be a net profit in any non-trivial case.  To assess this, let us
suppose that the pairing introduces a {\it momentum\/} gap of order
$\kappa$.  By this we mean that in an interval of order $\kappa$
around $p_F^A$ we will take superpositions of unoccupied and doubly
occupied states, as in ordinary BCS theory.  In particular, we do not
automatically fill the single-particle states for species B, even
though they are below the free-particle Fermi momentum $p_F^B$.  One
could also, more awkwardly but perhaps more properly, speak of 
normalized energy gaps of order $\kappa p_F^B /m_\al$ for the two
species.  However phrased, the point is that it is important to
prepare an equal number of modes to pair, and state-counting takes
place in momentum space.  The condensation energy must be of the same
energy as the spectral displacement of the particles, so we have for
the energy gain
$\epsilon_{\rm pair}$ per pair 
$\epsilon_{\rm pair} \sim {p_F^A \kappa /\tilde{m}}$, 
with $\tilde{m} \equiv m_A m_B/(m_A+m_B)$ the reduced mass.
On the other hand the density of pairs $n_{\rm pair}$ is of order
$n_{\rm pair} \sim 4\pi {p_F^A}^2 \kappa$.
In order to accommodate the depletion of species B, 
which is of the same order as the number of pairs, 
we must promote a corresponding number
of particles from $p_F^A$ to $p_F^B$, which costs 
$\frac{1}{2m_B} ({p_F^B}^2 - {p_F^A}^2)$
per pair.  Putting it all together, to make a net profit we require 
$
\epsilon_{\rm pair} \sim {p_F^A \kappa / \tilde{m}} 
> ({p_F^B}^2 - {p_F^A}^2 )/2m_B.
$

The region of pairing in momentum space will be strictly in
the interior of the larger Fermi surface --- clearly distinct from its
boundary ---  and we will realize interior gap superfluidity, if
$p_F^B - p_F^A > \kappa$. 
This is compatible with our earlier condition for
\begin{equation} 
1 > \frac{p_F^B + p_F^A}{2p_F^A} \frac{m_A}{m_A + m_B}. \label{eq:IGc}
\end{equation}
This consistency condition can be satisfied, specifically for 
$m_B \gg m_A$.  Note particularly that it is independent of the gap $\kappa$.
Looking back to our net profit condition, we see that $\kappa$ can be
taken arbitrarily small, for sufficiently large $m_B/m_A$.  
Thus interior gap superfluidity
can take place at
weak coupling,  where the mean-field
assumptions implicit in this heuristic analysis are valid.
In the class of models under discussion, therefore, we find a robust
weak-coupling phase characterized by a (momentum) 
gapped Fermi surface interior
to a surface with unpaired excitations.  For charged fermions, it is a
phase that is simultaneously superconducting and metallic at zero
temperature.

To construct the interior gap ground state explicitly we generalize
the standard BCS 
wavefunction as follows
$$
|\Psi_{\rm IG}\> =  \prod_{|\Vp|\leq p_\Delta} (\sin\theta_\Vp +
\cos\theta_\Vp \psi^\dag_{A\Vp}\psi^\dag_{B,-\Vp}) 
\prod_{|\Vp|>p_\Delta} \psi^\dag_{B\Vp} |0\>
$$
where the $\theta_\Vp$'s {\it and\/} $p_\Delta$ are variational
parameters.   
As usual, there is a manifold
of degenerate states featuring an overall relative phase between 
the $\sin\theta_\Vp$ and
$\cos\theta_\Vp$ factors. The order parameter is of usual form:
$\avg{\psi^\dag_{A\Vp}
\psi^\dag_{B,-\Vp}}_{\rm IG} =  \sin\theta_\Vp\cos\theta_\Vp$.
Upon variation with respect to $\theta_\Vp$  and $p_\Del$, 
we find
$\cos^2\theta_\Vp={1\over 2}\left(1 - {\eps^+_\Vp \over
\sqrt{(\eps^+_\Vp)^2 +\Del^2}}\right)$ and 
$p_\Del^2 = {1\over 2}({p^B_F}^2 + {p^A_F}^2) 
- {1\over 2} [-16 \Del^2 m_A m_B + ({p^B_F}^2 - {p^A_F}^2)^2]^{1/2} $, 
with $\eps^+_\Vp \equiv {1\over 2} (\eps^A_\Vp+\eps^B_\Vp)$. 
The gap parameter, defined here as 
$\Del= - g \sum_{|\Vp|\leq p_\Del} \avg{\psi^\dag_{A\Vp}
\psi^\dag_{B,-\Vp}}_{\rm IG} $, satisfies the integral equation
$1=-g\sum_{|\Vp|\leq p_\Del} {1\over \sqrt{{\eps^+_\Vp}^2 +\Del^2}} $.
 
An important departure from standard BCS theory occurs because the
energy difference between paired and unpaired modes no longer becomes
arbitrarily small. 
For this reason one is not doing degenerate perturbation theory, and
does not encounter a true infrared divergence (vanishing energy 
denominators) in the Cooper pairing channel.  As a consequence, the
gap equation supports a  non-zero solution only for $|g|> g_c$, with 
\be \textstyle
g_c \simeq {2 \over  N_+(0)\ln \left( {p_0 \, \lambda  \over
    {p_F^B}^2-{p_F^A}^2} {m_A+m_B\over m_A} \right)} \,, \label{eq:gc}
\ee
where we have introduced the generalized density of states
$N_+(0) \equiv \sum_{\Vp} \delta\big(\eps^+_\Vp\big)$ and $p_0$ is the
point where $\eps^+(\Vp_0)=0$.
Note however that $g_c\rightarrow 0$ when $m_B/m_A\rightarrow \infty$ for
fixed $p_F^{A,B}$,  so that interior gap superfluidity can be 
favorable for arbitrarily weak attractive
interactions, as we anticipated.   Numerically, we  typically find 
that $g_c$ is rather smaller. 

It is straightforward to calculate the condensation energy.  Up to
terms of higher order in $\Delta$, we find  
\be \textstyle
E_{\rm IG} -E_{\rm N}= -{1\over 2} N_+(0)\Del^2 \left[{1\over 2}
+  x_\Del e^{{\rm arcsinh}\,  x_\Del}
\right]
\ee
where 
$x_\Del={\eps^+(\Vp_\Del)/2\Del}$ 
and $E_N$ is the normal state energy
at $\Del=0$. For weak pairing $p_\Del$ is close to $p_F^A$, in which case
$\eps^+(p_\Del)$ is negative and finite. In the limit 
$\Del\rightarrow 0$, $x_\Del$ approaches  minus infinity, and the
two terms in the brackets  cancel. 
One sees that, for a given gap parameter $\Del$, 
the interior gap state gains  less condensation energy
than a conventional BCS state.

\begin{figure}[htbp]
\begin{center}
\includegraphics[width=\linewidth]{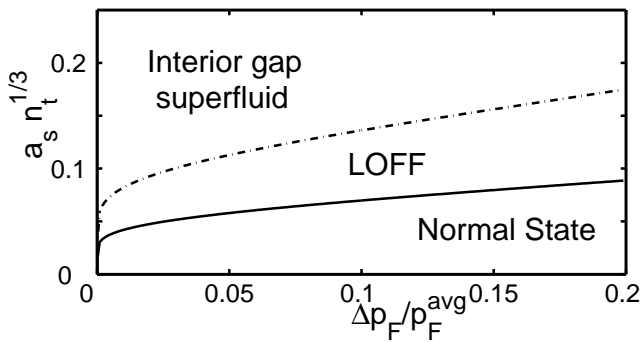}
\end{center}
\caption{A typical phase diagram, with the mass ratio $m_B/m_A=7$. 
Our coupling
constant  is $g= - 4\pi \hbar^2 a_s/\tilde{m}$, where $a_s$ is
the scattering length for $s$-wave scattering between species A and B, 
and $n_t=n^A+n^B$ is
the total density of particles, calculated using the 
formula $p_F^\al = (6\pi^2 n^\al)^{-1/3}$ 
as for a free Fermi gas. $\Del p_F\equiv p_F^B-p_F^A$ and $p_F^{\rm
avg} \equiv {1\over 2} (p_F^B+p_F^A)$.  
The dash-dotted line is plotted using the approximate analytical formula
(\ref{eq:gc}), while the solid line is plotted by adapting 
Eq. (3.12) of Takada and Izuyama \cite{Takada+:69} 
to the case at hand.   Were it plotted directly from our numerical
simulations, the phase transition line 
between interior gap and LOFF  would be slightly lowered.}
\label{fig:phase}
\end{figure}

Thus we have demonstrated that the normal state is 
unstable against formation of an interior gap superfluid  
when the attractive coupling is strong enough, i.e., $|g|> g_c$. 
The LOFF state is another candidate 
for pairing of mismatched  Fermi surfaces. We have calculated the ground-state 
energy difference between these two candidate states numerically. 
Fig.~\ref{fig:phase} shows a
typical phase diagram.
We have considered here only the simplest LOFF state, with
pairing order parameter $\propto \Del_{\rm LOFF} e^{i\VQ\cdot \Vx}$
with $|\VQ|$ of order 
$p_F^B-p_F^A$ \cite{LOFF:65+64,Takada+:69}.   

A few remarks concerning this phase diagram are in order.
The critical coupling constant $g_c$ in
Eq.~(\ref{eq:gc}) in principle governs  
the (second-order) phase transition  from the normal state
to the interior gap state, rather than that from the LOFF state. 
However, we find in our numerical calculation that the phase transition
line between the interior gap and LOFF states roughly coincides with
the onset of the interior gap ordering.  Indeed, since an LOFF particle
pair carries a total momentum $\VQ\neq 0$, the pairing process in the
LOFF state occurs mostly within the intersection of two closed shells
of thickness $\kappa$, one centered on $\VQ$ and the other on
zero. The intersection of these shells is a closed ring of thickness
$\kappa$. By contrast, for the interior gap state pairing occurs
within a full two-dimensional shell.  Since the density of states
involved in pairing for the interior gap state is larger than that for
the LOFF state, the former develops an order parameter which increases
(as a function of intrinsic coupling strength) exponentially faster
than the latter, and rapidly dominates once it sets in.  On the other
hand, the LOFF phase can be realized also for $m_A > m_B$, that is
when the heavier fermion has the smaller Fermi surface, when the
interior gap phase is not available.  This is the case of primary
interest for possible phases of quark matter in neutron star
interiors, where the B species is the strange quarks.
BCS states
correspond to the region in our phase diagram where the two species have
approximately equal Fermi momentum, $\Del p_F \sim 0$.

\begin{figure}[htbp]
\begin{center}
\includegraphics[width=\linewidth]{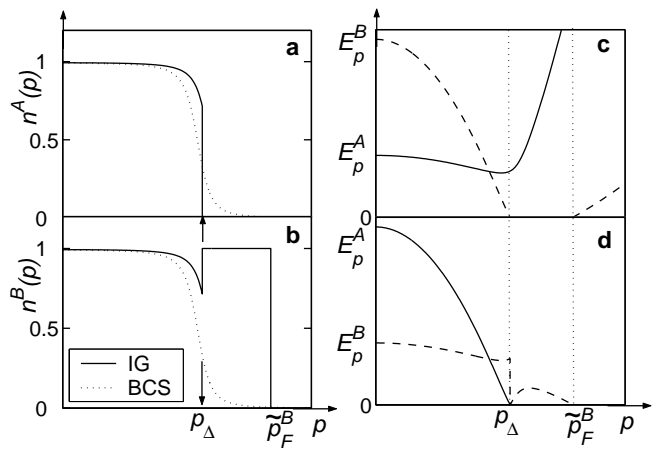}
\end{center}
\caption{
{\bf a} and {\bf b:}  A plot of 
average particle occupation number 
$n^{A}_\Vp$  and $n^{B}_\Vp$, respectively, 
in the interior gap superfluid state in comparison with those in a
conventional BCS state. For the comparison, we assume the same gap parameter  
and a matched Fermi surface at $p=p_F^A$.  
{\bf c.} Quasiparticle energy spectra to add a species A or B particle
to the system.
Corresponding  quasi-particle states are 
$\psi^\dag_{A\Vp} | \Psi_{\rm IG}\>$ (`---' line) and 
$\psi^\dag_{B\Vp} | \Psi_{\rm  IG}\>$ (`- - -' line). 
{\bf d.} Quasiparticle energy spectra to remove a species A or B particle
from the system.
Corresponding  quasi-particle states are 
$\psi_{A\Vp} | \Psi_{\rm IG}\>$ (`---' line) and 
$\psi_{B\Vp} | \Psi_{\rm  IG}\>$ (`- - -' line). 
}
\label{fig:qpnE}
\end{figure}

The average particle occupation number
$n^{A,B}_\Vp$ has an unconventional form. $n^A_\Vp=n^B_\Vp =
\cos^2\theta_\Vp$ for $|\Vp| \leq p_\Del$, and $n^A_\Vp=0$ and $n^B_\Vp
=1$ for $p_\Del< |\Vp| \leq \tilde{p}_F^B$, where $\tilde{p}_F^B$ is
the shifted Fermi momentum of species B particle due to pairing interaction:
$\tilde{p}_F^B \simeq p_F^B(1+ p_0\tilde{m} \Del e^{{\rm arcsinh}
x_\Del}/{p_F^B}^3)$. 
Fig.~\ref{fig:qpnE}
shows the average particle occupation number as a
function of momentum at zero temperature.
In the interior gap state, the pairing correlation smears 
the  species A Fermi surface slightly ---  a fraction of
single-particle states are depleted below
$p_F^A$  and inserted back between the normal state Fermi surface at
$p_F^A$ and the maximum pairing surface at $p_\Del$.
This distribution does not differ qualitatively from
what one finds in a conventional BCS superconducting state. 
Species B displays a more dramatic contrast. Some modes that are occupied
in the normal state, in the Fermi ball interior around $p=p_F^A$,
are now depleted, and the deficit is made up at the top of the 
Fermi surface.  This
enlarges the Fermi surface from $p_F^B$ to
$\tilde{p}_F^B$.   Discontinuities in the distribution 
$n^B_\Vp$ occurs at both $p_\Del$ and $\tilde{p}_F^B$.

Approximate quasi-particle excitations of the interior gap state 
are obtained by diagonalizing  the Hamiltonian (\ref{eq:H}) at 
mean-field level using the Bogoliubov transformation
$\gamma_{1\Vp}^\dag = \sin\theta_\Vp \psi^\dag_{A\Vp} -\cos\theta_\Vp
\psi_{B,-\Vp}$
and   
$\gamma^\dag_{2,-\Vp} = \sin\theta_\Vp \psi^\dag_{B,-\Vp} + \cos\theta_\Vp
\psi_{A\Vp}$. These operators create quasiparticle
excitations $\gamma^\dag_{1,2;\Vp} |\Psi_{\rm IG}\>$ from the interior
gap superfluid state. 
Their spectra are given by
\be \textstyle
E_{1,2}(\Vp) = {1\over 2}(\eps^A_\Vp-\eps^B_\Vp) \pm \sqrt{{\eps^+_\Vp}^2
  +\Del^2} \,,   \label{eq:E12}
\ee
with  
all energies measured from the shifted
Fermi surfaces 
defined by $\tilde{p}_F^A\simeq p_F^A$ and 
$\tilde{p}_F^B$. 
Unlike in conventional BCS theory, there are two branches of
excitations: $E_1(\Vp)$ is gapped while $E_2(\Vp)$ is gapless at both
$|\Vp|=p_\Del$ and $|\Vp|=\tilde{p}^F_B$. 
Notably, 
`$-E_{2\Vp}$'  becomes negative for  momenta 
$|\Vp|$ between $p_\Del$ and
$\tilde{p}_F^B$. This is interpreted consistently as indicating that
all such states are 
filled by species-B particles (note $\sin\theta_\Vp =0$), and that the
absolute value   
$|E_{2\Vp}|$ is the energy to create  a hole excitation there
(Fig.~\ref{fig:qpnE}).

Given the explicit form of the quasiparticles and their spectrum,
phenomenological consequences can be derived along standard lines.
The novelty of the interior gap state is that a large manifold of
low-energy ``normal state'' excitations coexists with superfluidity.
This spectrum could be probed directly in tunneling experiments.  At
finite temperature the normal state excitations will be excited, and
the appropriate description will involve a two-fluid model incorporating
dissipation.  In these regards there is some resemblance to
conventional superfluids whose order parameter has nodes, such as the
$p$-wave superfluid in liquid $^3$He, or the $d$-wave superconducting
cuprates.  But these states differ from interior gap superfluids both
quantitatively, in that the density of gapless modes is much smaller,
and qualitatively, in that they involve breaking of rotational
symmetry.  Another partial analogue is the Abrikosov-Gor'kov gapless
superconductivity with magnetic impurities \cite{Abrikosov:61}; but of
course here we do have a gap, and impurities are not a central issue.

Interior gap superfluidity will be realized in a two-species mixture
of fermionic cold atoms with different mass. Recent theoretical
\cite{Stoof+:96,+Stoof+:97,Holland+:01,+Griffin:02pre,Hofstetter+:02pre,Combescot:01}
and experimental \cite{+Jin:99,+Jin:01,+Thomas:00,+Thomas:02} efforts
point to the possibility of 
superfluidity  in two-state mixtures of $^6$Li or $^{40}$K atoms. 
A stable mixture with different mass could 
be realized, for instance, in the $^6$Li and $^{40}$K atomic gas
\cite{Ketterle:com}.  Despite its qualitative difference from BCS-type
states, the interior gap state is estimated to have a superfluid
transition temperature of the same order as that of
Refs.~\cite{Stoof+:96,+Stoof+:97,Holland+:01,+Griffin:02pre,Hofstetter+:02pre,Combescot:01},
for not too weak coupling.


Also, we perceive no problem of principle forbidding the realization
of interior gap superfluidity in electron gases, where the species are
electrons from different bands, which can have markedly different
effective masses.  
The case of electrons coming from two bands differs however 
from the atomic case in
that one should specify a
single density and an energy offset between band minima,
instead of two independent densities (or
chemical potentials).  There can be phase transitions as a function of the
single density, for example
between interior gap and conventional BCS-type order.   Also, of course,
the dispersion relations can be
different from simple parabolas, which has interesting consequences.  We
shall return
to these questions in a future publication. 

We have benefited from discussions with J. Bowers, C.
Honerkamp, W. Ketterle, K. Rajagopal, and X.-G. Wen.
This work is supported in part by funds provided by the
U.S. Department of Energy (D.O.E.) under cooperative research
agreement \#DF-FC02-94ER40818.


\bibliography{interior_gap}


\end{document}